\documentclass[twocolumn, pra, aps, nofootinbib, superscriptaddress]{revtex4-1}
\usepackage[title]{appendix}
\usepackage{amsmath}
\usepackage{amssymb}
\usepackage{physics}
\usepackage[dvips]{graphicx}
\usepackage[normalem]{ulem}
\usepackage{comment}
\usepackage{algpseudocode,algorithm}
\usepackage{caption}
\usepackage{subcaption}
\usepackage{qcircuit}
\usepackage{bbm}

\usepackage{amsthm}
\usepackage{hyperref}
\hypersetup{
  colorlinks   = true, %Colours links instead of ugly boxes
  urlcolor     = blue, %Colour for external hyperlinks
  linkcolor    = blue, %Colour of internal links
  citecolor   = blue%Colour of citations
}

\usepackage{tikz}
\usepackage{multirow}
\usepackage{enumitem}
\usepackage{url}

\makeatletter
\newtheorem*{rep@theorem}{\rep@title}
\newcommand{\newreptheorem}[2]{%
\newenvironment{rep#1}[1]{%
 \def\rep@title{#2 \ref{##1}}%
 \begin{rep@theorem}}%
 {\end{rep@theorem}}}
\makeatother

\begin{document}
\title[Short Title]{Timing and resource-aware mapping of quantum circuits to superconducting processors}

\author{Lingling Lao}
\affiliation{QuTech, Delft University of Technology, The Netherlands}
\affiliation{Department of Quantum and Computer Engineering, Delft University of Technology, The Netherlands}
\affiliation{Department of Physics and Astronomy, University College London, UK}
\author{Hans van Someren}
\author{Imran Ashraf}
\author{Carmen G. Almudever}
\affiliation{QuTech, Delft University of Technology, The Netherlands}
\affiliation{Department of Quantum and Computer Engineering, Delft University of Technology, The Netherlands}

\begin{abstract}
Quantum algorithms need to be compiled to respect the constraints imposed by quantum processors, which is known as the mapping problem. 
The mapping procedure will result in an increase of the number of gates and of the circuit latency, decreasing the algorithm's success rate. 
It is crucial to minimize mapping overhead, especially for Noisy Intermediate-Scale Quantum (NISQ) processors that have relatively short qubit coherence times and high gate error rates.
Most of prior mapping algorithms have only considered constraints such as the primitive gate set and qubit connectivity, but the actual gate duration and the restrictions imposed by the use of shared classical control electronics have not been taken into account. 
In this paper, we present a timing and resource-aware mapper called Qmap to make quantum circuits executable on a scalable superconducting processor named Surface-17 with the objective of achieving the shortest circuit latency.
In particular, we propose an approach to formulate the classical control restrictions as resource constraints in a conventional list scheduler with polynomial complexity.
Furthermore, we implement a routing heuristic to cope with the connectivity limitation.
This router finds a set of movement operations that minimally extends circuit latency.
To analyze the mapping overhead and evaluate the performance of different mappers, we map 56 quantum benchmarks onto Surface-17.
Compared to a prior mapping strategy that minimizes the number of operations, Qmap can reduce the latency overhead up to $47.3\%$ and operation overhead up to $28.6\%$, respectively.
\end{abstract}

\maketitle

\section{Introduction}
\label{sec:intro}
Quantum computing is entering the Noisy Intermediate-Scale Quantum (NISQ) era \cite{preskill2018quantum}. 
This refers to exploiting quantum processors consisting of only 50 to a few hundreds of noisy qubits - i.e. qubits with a relatively short coherence time and faulty operations. 
Due to the limited number of qubits, hardly or no quantum error correction (QEC) will be used in the next coming years, posing a limitation on the size of the quantum applications that can be successfully run on NISQ processors. 
Nevertheless, these processors will still be useful to explore quantum physics, and implement small quantum algorithms that will hopefully demonstrate quantum advantage~\cite{feynman1982simulating}.
For running quantum applications on NISQ devices, it is thus crucial to minimize their size in terms of circuit width (number of qubits), number of gates, and circuit latency/depth (number of cycles/steps).
In addition, these quantum applications have to be adapted to the hardware constraints imposed by quantum processors. 
The main constraints include:

\begin{itemize}
    \item \textbf{Primitive gate set:} Generally, only a limited set of quantum gates that can be realized with relatively high fidelity will be predefined on a quantum device.
    Each quantum technology may support a specific universal set of single-qubit and two-qubit gates, which are called primitive gates.
    Different primitive gates may have different gate durations.
    For instance, some superconducting quantum technologies have CZ as a primitive two-qubit gate of which the duration is twice as long as of a single-qubit primitive gate~\cite{kjaergaard2019superconducting}.
    
    \item \textbf{Qubit connectivity:} quantum technologies such as superconducting qubits~\cite{ibm17experience,boixo2018characterizing,rigetti} and quantum dots~\cite{hill2015surface, li2018crossbar} arrange their qubits in 1D/2D architectures with \textit{nearest-neighbour} (NN) interactions. 
    This means that only neighbouring qubits can interact or in other words, qubits are required to be adjacent for performing a two-qubit gate. 
    
    \item \textbf{Classical control:} classical electronics are required for controlling and operating the qubits. 
    Using a dedicated instrument per qubit is not scalable and is a very expensive approach. 
    Therefore, shared control is required especially when building scalable quantum processors. 
    For instance, a single Arbitrary Waveform Generator (AWG) is used for operating on a group of qubits and several qubits are measured through the same feedline  \cite{asaad2016independent, mckay2018qiskit}. 
\end{itemize}

All these constraints may vary between different quantum processors, and quantum circuits normally cannot be directly executable on these devices.
A mapping procedure is required to transform a hardware-agnostic quantum circuit into a constraint-compliant one that can be realized on a given device. This mapping process i) decomposes any quantum gate into the supported primitive gates; ii) performs an initial placement of qubits and finds the set of movement operations to route non-NN qubits to adjacent positions when they need to interact; and iii) schedules operations to leverage the maximum available parallelism.
Moreover, minimizing mapping overhead in terms of the number of gates and circuit execution time (latency) is critical for implementing quantum algorithms on NISQ processors.

Different solutions including both exact algorithms and heuristics have been proposed to map quantum circuits onto NISQ processors.
\cite{yazdani2013quantum,  lye2015determining, wille2016look,farghadan2017quantum, herbert2018using} propose mapping approaches for a 2D grid qubit architecture with NN interactions.
Other works ~\cite{qiskit, Zulehner2018efficient, siraichi2018qubit, finigan2018qubit, li2019tackling, tannu2019not, nishio2019extracting, cowtan2019qubit,rigetti,venturelli2018compiling, booth2018comparing, venturelli2019quantum} target current quantum processors from IBM and Rigetti which have irregular qubit connections.
Most of prior works~\cite{ yazdani2013quantum,  lye2015determining, wille2016look,farghadan2017quantum, herbert2018using, qiskit, Zulehner2018efficient, siraichi2018qubit, finigan2018qubit, li2019tackling, tannu2019not, nishio2019extracting, cowtan2019qubit,rigetti} mainly consider the qubit connectivity and the primitive gate set constraints and their strategies focus on minimizing gate overhead.
They assume that any operation takes one time-step without taking the actual gate duration into account.
Moreover, they do not consider the shared classical control electronics, which restricts the parallelism of some operations.
This means the output circuits from previous mapping algorithms need to be further scheduled by another hardware-aware translation phase such as OpenPulse from IBM~\cite{mckay2018qiskit} so that quantum operations can be performed on real qubits with correct timing without violating any classical control constraint~\cite{mckay2018qiskit, versluis2017scalable}.
Venturelli et al.
~\cite{venturelli2018compiling, booth2018comparing, venturelli2019quantum} consider gate duration and crosstalk constraints, but their mathematical optimization formulation has exponential complexity.

This paper presents a timing and resource-aware mapper called \textbf{Qmap} to make quantum circuits executable on the Surface-17 superconducting processor~\cite{versluis2017scalable}.
Different modules are developed in Qmap to comply with the hardware constraints, including common restrictions such as primitive gate set and qubit connectivity, as well as other hardware parameters such as actual gate duration and classical control constraints which have not been addressed in prior works. 
Qmap is embedded in the OpenQL compiler
 \cite{openql18qutech} and its output circuit is described by an
executable low-level QASM-like code with precise timing information.
In order to analyze the impact of the mapping procedure, we compile 56 benchmarks taken from RevLib~\cite{wille2008revlib} and QLib~\cite{lin2014qlib} onto the Surface-17 processor. 
Compared to the original circuit characteristics before mapping,
the evaluation results show that the circuit latency and the number of operations after mapping can increase up to $260\%$ and $78.1\%$, respectively.
   
The main contributions of this paper are the following:

\begin{itemize}
    \item We provide a comprehensive analysis of the hardware constraints of the Surface-17 processor, including the supported primitive gates with corresponding duration, the processor's topology that limits the qubit connectivity, and the classical control constraints resulting from the shared control electronics among qubits.
    
    \item We develop a Qmap mapper embedded in the OpenQL compiler~\cite{openql18qutech} to compile a quantum circuit into one that complies with all the above constraints of Surface-17.
    Specifically, we propose an approach to formulate the classical control limitations as resource constraints in a conventional list scheduling algorithm.
    Its objective is to achieve the shortest circuit latency and therefore the highest gate-level parallelism with respect to these constraints.
    The complexity of the developed scheduling heuristic is polynomial in terms of the number of operations and resources, which is applicable to large-scale circuits.
    
    \item For coping with the limited qubit connectivity, we present a routing strategy in Qmap to move qubits that need to interact to be adjacent.  
    The proposed router not only finds shortest paths that use least number of operations for moving qubits (which is the routing strategy developed in prior works) but also selects a set of movement operations that will minimally extend the overall circuit latency.
    Compared to a prior mapping strategy, the average reduction of latency overhead and the average reduction of gate overhead when using Qmap are $22\%$ and $3.0\%$, respectively.
    
    \item To enable a flexible implementation, we provide a method to encode all hardware characteristics in a configuration file that is accessed by every module of the compiler.
    This flexibility also allows a comparative analysis of the mapping impacts of different characteristics, giving some directions for building future quantum devices. 
    In addition, it allows the mapper to target different processors. 
    
    \item Qmap uses not only SWAP operations (3 consecutive CNOTs) for moving qubits but also MOVE operations (2 consecutive CNOTs) when possible.
    Compared to the mapping by only using SWAPs in prior works, the use of MOVEs helps to reduce the number of gates and the circuit latency up to $38.9\%$ and $29\%$ respectivel.
    
\end{itemize}

The rest of this paper is organized as follows.
We first describe all the hardware parameters that will be considered in this work in Section \ref{sec:constraints}.
Then we introduce the proposed resource-constrained scheduling algorithm in Section~\ref{sec:schedule} and other modules of the developed mapper such as the routing heuristic in Section \ref{sec:mapper}. 
Afterwards, we evaluate this mapping strategy in Section \ref{sec:results} and summarize related works in Section \ref{sec:review}. 
Finally, Section \ref{sec:conclusion} concludes the paper and discusses future work.

\section{Quantum hardware constraints}
\label{sec:constraints}
In this section, the hardware constraints of the Surface-17 superconducting processor will be briefly introduced, including the primitive gates that can be directly performed, the topology of the processor which limits interactions between qubits, and the constraints caused by the classical control electronics which impose extra limitations on the parallelism of the operations. 

\subsection{Primitive gate set}

In order to run any quantum circuit, a universal set of operations needs to be implemented.  
In superconducting quantum processors, these operations commonly are measurement, single-qubit rotations, and multi-qubit gates.

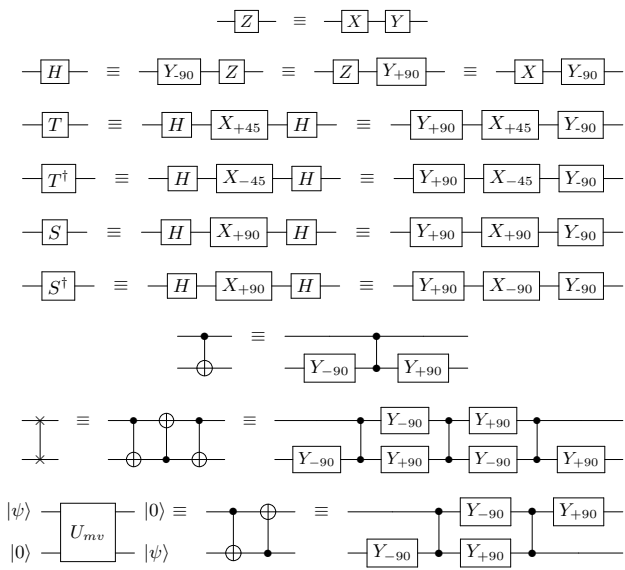
\begin{figure}[tbh!]   
\begin{center}
\resizebox{0.16\textwidth}{!}{
\Qcircuit @C=1em @R=.7em {
& \gate{Z} & \qw & \push{\equiv} &  & \gate{X} & \gate{Y} & \qw \\
}}\vspace{3mm}
\resizebox{0.45\textwidth}{!}{
\Qcircuit @C=1em @R=.7em {
    & \gate{H} & \qw & \push{\equiv} & & \gate{Y_{\text{-90}}} & \gate{Z} & \qw & \push{\equiv} & & \gate{Z} & \gate{Y_{\text{+90}}} & \qw & \push{\equiv} & & \gate{X} & \gate{Y_{\text{-90}}} & \qw\\
}}\vspace{3mm}
\resizebox{0.45\textwidth}{!}{
\Qcircuit @C=1em @R=.7em {
& \gate{T} & \qw & \push{\equiv} &  & \gate{H} & \gate{X_{+45}} & \gate{H} & \qw & \push{\equiv} &  & \gate{Y_{\text{+90}}} & \gate{X_{+45}} & \gate{Y_{\text{-90}}} & \qw \\
}}\vspace{3mm}
\resizebox{0.45\textwidth}{!}{
\Qcircuit @C=1em @R=.7em {
& \gate{T^{\dagger}} & \qw & \push{\equiv} &  & \gate{H} & \gate{X_{-45}} & \gate{H} & \qw & \push{\equiv} &  & \gate{Y_{\text{+90}}} & \gate{X_{-45}} & \gate{Y_{\text{-90}}} & \qw \\
}}\vspace{3mm}
\resizebox{0.45\textwidth}{!}{
\Qcircuit @C=1em @R=.7em {
& \gate{S} & \qw & \push{\equiv} &  & \gate{H} & \gate{X_{+90}} & \gate{H} & \qw & \push{\equiv} &  & \gate{Y_{\text{+90}}} & \gate{X_{+90}} & \gate{Y_{\text{-90}}} & \qw \\
}}\vspace{3mm}
\resizebox{0.45\textwidth}{!}{
\Qcircuit @C=1em @R=.7em {
& \gate{S^\dagger} & \qw & \push{\equiv} &  & \gate{H} & \gate{X_{+90}} & \gate{H} & \qw & \push{\equiv} &  & \gate{Y_{\text{+90}}} & \gate{X_{-90}} & \gate{Y_{\text{-90}}} & \qw \\
}}\vspace{3mm}
\resizebox{0.22\textwidth}{!}{
\Qcircuit @C=1em @R=.7em {
 & \ctrl{1} & \qw & \push{\equiv} &  & \qw & \ctrl{1} & \qw & \qw \\
 & \targ & \qw &  &  & \gate{Y_{-90}} & \control \qw & \gate{Y_{+90}} & \qw\\
 }}\vspace{3mm}
\resizebox{0.45\textwidth}{!}{
\Qcircuit @C=1em @R=.7em {
 & \qswap & \qw  & \push{\equiv}& &\ctrl{1} &\targ & \ctrl{1}&\qw &\push{\equiv}& &\qw&\ctrl{1}&\gate{Y_{-90}}&\control \qw&\gate{Y_{+90}}&\ctrl{1}&\qw&\qw\\
 & \qswap \qwx& \qw & & & \targ &\ctrl{-1}& \targ&\qw& & & \gate{Y_{-90}} &\control \qw&\gate{Y_{+90}}&\ctrl{-1}&\gate{Y_{-90}} &\control \qw&\gate{Y_{+90}}& \qw\\
 }}\vspace{3mm}
 \resizebox{0.45\textwidth}{!}{
\Qcircuit @C=1em @R=.7em {
 &\lstick{\ket{\psi}}& \multigate{1}{U_{mv}} &\rstick{\ket{0}} \qw  & &\push{\equiv}& &\ctrl{1} &\targ & \qw &\push{\equiv}& &\qw&\ctrl{1}&\gate{Y_{-90}}&\control \qw&\gate{Y_{+90}}&\qw\\
 &\lstick{\ket{0}} & \ghost{U_{mv}}& \rstick{\ket{\psi}}\qw & & & & \targ &\ctrl{-1}& \qw& & & \gate{Y_{-90}} &\control \qw&\gate{Y_{+90}}&\ctrl{-1}&\qw&\qw\\
 }}
\end{center}
\caption{Gate decomposition into primitives supported in the superconducting Surface-17 processor. $U_{mv}$ is the MOVE operation. }
\label{decompositions}
\end{figure}

In principle, any kind of single-qubit rotation can be performed on the Surface-17 processor.
However, an infinite  amount of gates cannot be predefined. In this work, we will limit single qubit gates to X and Y rotations (easier to implement), and more specifically  $\pm$ 45, $\pm$ 90 and $\pm$ 180 degrees will be used in our decomposition. The primitive two-qubit gate on this processor is the conditional-phase (CZ) gate.
Table~\ref{tbl:primitive} shows the gate duration (gate execution time) of single-qubit gates, CZ gate and measurement (in the Z basis)~\cite{brien2017density}.
After mapping, the output circuit will only contain
operations that belong to this primitive gate set.
The decomposition for $Z, H, S, S^{\dagger}, T, T^{\dagger}$, CNOT, SWAP and MOVE gates into these primitive gates is shown in Figure~\ref{decompositions} (ignoring the global phase).

\begin{table}[bth!]
\centering
\caption{The gate duration in cycles (each cycle represent 20 nanoseconds) of the primitive gates in the Surface-17 processor.}
\label{tbl:primitive}
\begin{tabular}{c|c}
\hline
\textbf{Gate type} & \textbf{Duration}      \\ \hline
$R_{X}(\pm 45, \pm 90, \pm180)$  & 1 cycle  \\ 
$R_{Y}(\pm 45, \pm 90, \pm180)$  & 1 cycle  \\ 
CZ        & 2 cycles   \\ 
$M_{Z}$         &  15 cycles    \\\hline
\end{tabular}
\end{table}

\subsection{Processor topology}
\label{sec:topology}

\begin{figure}[htb!]
\centering
\includegraphics[width=0.22\textwidth]{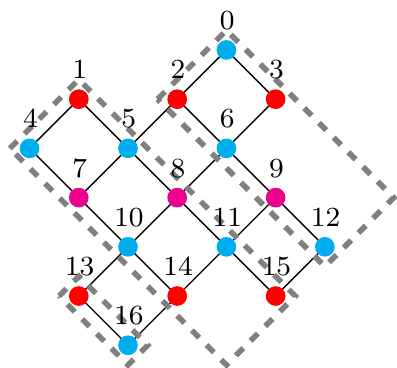}
\caption{Schematic of the realization of Surface-17 superconducting processor.}
\label{sc17}
\end{figure}

Figure \ref{sc17} shows the topology of the Surface-17 processor, where nodes represent the qubits and edges represent the connections (resonators) between them.
Two-qubit gates can only be performed between connected qubits, i.e., \textit{nearest-neighbouring} qubits.
This implies that qubits that have to interact but are not placed in neighbouring positions will need to be moved to be adjacent. 
Quantum states in superconducting technology are usually moved using SWAP gates. 
A SWAP gate is implemented by three CNOTs that in the case of the Surface-17 processor need to be further decomposed into CZ and $R_Y$ gates as shown in Figure \ref{decompositions}. 
In this work, we also consider the use of a MOVE operation which only requires two CNOTs (see Figure \ref{decompositions}). 
Note that a MOVE operation requires that the destination qubit where the quantum state needs to be moved to, is in the $\ket{0}$ state. 
As mentioned, moving qubits results in an overhead in terms of number of operations and circuit depth, which in turn will decrease the circuit reliability.
Therefore, an efficient routing procedure is required to find the series of movement operations to enable all two-qubit gates with minimum overhead.

\subsection{Classical control constraints}
In principle, any qubit in a processor can be operated individually and then any combination of independent single-qubit and two-qubit operations can be performed in parallel. 
However, scalable quantum processors use classical control electronics with channels that are shared among several qubits. 
Here we will describe the constraints imposed by the classical control electronics used in the Surface-17 processor and how they affect the parallelism of quantum operations.

\paragraph{Single-qubit gates:} 
Single-qubit gates on superconducting qubits are performed by using microwave pulses. In Surface-17,
these pulses are applied at a few fixed specific frequencies to ensure scalability and precise control. The three frequencies used in Surface-17 are shown in Figure \ref{sc17}: single-qubit gates on red, blue and pink colored qubits are performed at frequencies $f_1$, $f_2$, and $f_3$, respectively~\cite{versluis2017scalable}. In this work, we assume that same-frequency qubits are operated by the same microwave source or arbitrary waveform generator (AWG) and a vector switch matrix (VSM) is used for distributing the control pulses modulated on the waves to the corresponding qubits \cite{asaad2016independent}. 

The consequence of this is that one can perform the same single-qubit gate on all or some of the qubits that share a frequency, but one cannot perform different single-qubit gates at the same time on these qubits (as these would require other pulses to be generated). For instance, an $X$ gate can be performed simultaneously on any of the pink qubits (7, 8 and 9) but not an $X$ and a $Y$ operation. 

\paragraph{Measurement:}
Measuring the qubits is done by using feedlines each of which is coupled to multiple qubits~\cite{versluis2017scalable}.
In Figure \ref{sc17}, qubits in the same dashed rectangle are using the same feedline, e.g., qubits 13 and 16 will be measured through the same feedline.
Because measurement takes several steps in sequence, measurement of a qubit cannot start when another qubit coupled to the same feedline is being measured, but any combination of qubits that are coupled to the same feedline can be measured simultaneously at a given time. For instance, qubits 13 and 16 can be measured at time $t_0$, but it is not possible to start measuring qubit 13 at time $t_0$ and then measure qubit 16 at time $t_1$ if the previous measurement has not finished. 

\paragraph{Two-qubit gates:}
As mentioned, in the processor of Figure~\ref{sc17} each qubit belongs to one of three frequency groups $f_1$ $>$ $f_2$ $>$ $f_3$, colored red, blue and pink, respectively; links between neighbouring qubits are either between qubits from $f_1$ and $f_2$, or between qubits from $f_2$ and $f_3$, i.e. between a higher frequency qubit and a next lower one. 
In between additional frequencies are defined: $f_1$ $>$ $f^{int}_1$ $>$ $f_2$ $>$ $f^{park}_2$ $>$ $f^{int}_2$ $>$ $f_3$ $>$ $f^{park}_3$ (see the frequency arrangement and the example interactions presented in Figure 5 of~\cite{versluis2017scalable}); each qubit can be individually driven with one of the frequencies of its group (e.g. group $\{f_{2},f^{int}_{2},f^{park}_{2}\}$). 
A CZ gate between two neighbouring qubits is realized by lowering the frequency of the higher frequency qubit near to the frequency of the lower one. 
For instance, a CZ gate between qubits 3 and 0 is performed by detuning qubit 3 from  $f_1$ to $f^{int}_1$, which is near to the frequency $f_2$ of qubit 0. 
However, CZ gates will occur between any two neighbouring qubits which have close frequencies and share a connection.
For example, a CZ gate can occur between the detuned qubit 3 in $f^{int}_1$ and its neighbour qubit 6 in $f_2$ in the above example. 
To avoid this, the qubits that should not be involved in a CZ gate must be kept out of the way. 
In this example, q6 needs to be detuned to a lower \textit{parking frequency}, $f^{park}_2$ . 
Note that, qubits in parking frequencies cannot engage in any two-qubit or single-qubit gate. 
In addition, when performing a CZ on qubits 3 and 0, qubit 2 must stay at $f_1$ (and not be detuned) to avoid interaction between qubits 2 and 0. 
The implementation of two-qubit gates poses limitations not only on parallelizing multiple two-qubit gates but also on the parallelism of two-qubit gates and single-qubit gates. More details can be found in~\cite{versluis2017scalable}. 

Violation of these classical control constraints will cause incorrect execution of quantum operations, leading to a computational failure.
Therefore, scheduling algorithms that can take these constraints into account are needed to explore the maximum available parallelism.

\subsection{Configuration file}
\label{sec:configuration}

The hardware characteristics explained in this section are precisely described in a configuration file (in json format).
It parameterizes the mapping modules that will be introduced in the next section.

\paragraph{Primitive gate set:}
For Surface-17, the primitive gates with all attributes including duration as listed in Table~\ref{tbl:primitive} and the gate decomposition rules corresponding to those in Figure~\ref{decompositions} are described in full detail in the configuration file.

\paragraph{Processor topology:} 
The topology is defined by describing each connection with its source and target qubits.
In Surface-17, all edges are bidirectional, e.g., both CNOT$(q_{a}, q_{b})$ and CNOT$(q_{b}, q_{a})$ can be performed on edge $e(q_{a}, q_{b})$.
Qubits and directed qubit connections are both named by integer values taken from contiguous ranges of integer numbers starting from 0. As an example, the qubit numbering of the Surface-17 processor is shown in Figure~\ref{sc17}; in the Surface-17 topology the number of directed qubit connections is $48$.

\paragraph{Classical control constraints:}
For \textbf{single-qubit gates}, we use a look-up table $T_{g1}$ to describe the available AWGs and the list of corresponding qubits that each AWG controls. 
Similarly for \textbf{measurement}, the feedlines (three feedlines in Surface-17) and the corresponding qubits that each feedline is coupled to are described in a look-up table $T_{gm}$ in the configuration file. 
The AWGs and feedlines are both named by contiguous integer numbers starting from 0.
As mentioned in Section~\ref{sec:constraints}, it is assumed that three AWGs and three feedlines are used in Surface-17, that is, $\left | T_{g1} \right |=3$ and $\left | T_{gm} \right |=3$, respectively.
The classical control constraints of \textbf{two-qubit gates} are defined by using two look-up tables. 
One called $T_{g2f}$ describes for each connection which other connections cannot be used to execute CZ gates in parallel (24 bi-directional edges on the Surface-17 topology, i.e. $\left | T_{g2f} \right |=48$).
The other table $T_{g2d}$ describes for each connection which set of qubits needs to be detuned in addition to one of its end-points, which means a CZ on this connection and single-qubit gates on these detuned qubits cannot be performed in parallel( $\left | T_{g2d} \right |=48$).

\section{Resource-constrained scheduling}
\label{sec:schedule}
Some current quantum technologies such as superconducting qubits and quantum dots have relatively short coherence times, limiting the size of circuits that can be run successfully with high fidelity. 
It is therefore necessary to minimize the execution time of the circuit (or $makespan$, or circuit latency) and explore the highest gate-level parallelism, which is the objective of a quantum gate scheduler. 
Before discussing the other mapping modules, we first introduce the proposed heuristic scheduling algorithm that can take the actual gate duration and classical control constraints into account.
% This since it is the base of both the router and the global scheduler.
The circuit shown in Figure~\ref{circuit_qasm} will be used as an example.
We refer to the qubits in the quantum circuit as virtual qubits (others call them program qubits or logical qubits). 
These need to be mapped to the qubits in the quantum processor called physical, real or hardware qubits or locations

\subsection{Weighted dependency graph}
As mentioned previously, precise timing is essential for correctly executing quantum applications on real qubits. 
Therefore, a scheduler that considers gate duration is required to efficiently generate the correct instruction sequences with timing information meanwhile minimizing the circuit execution time.
Prior works \cite{yazdani2013quantum,  lye2015determining, wille2016look,farghadan2017quantum, herbert2018using, qiskit, Zulehner2018efficient, siraichi2018qubit, finigan2018qubit, li2019tackling, tannu2019not, nishio2019extracting,cowtan2019qubit, rigetti} do not consider the actual gate duration, assuming any operation takes one time-step.
To ensure quantum operations can be executed at correct time, their output circuits need to be further scheduled by some other low-level hardware-aware units such as OpenPulse~\cite{mckay2018qiskit}.
In contrast, the scheduling algorithm developed in the Qmap mapper will directly take gate duration into account. 

\begin{figure}[htb!]
\centering
	\begin{subfigure}[b]{0.2\textwidth}
		\includegraphics[height=1\textwidth]{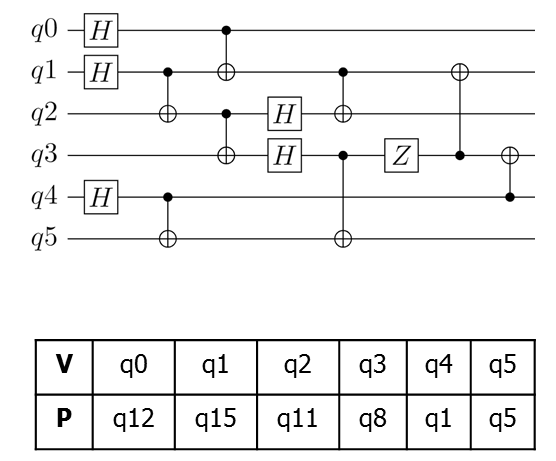}
		\caption{}
		\label{circuit}
	\end{subfigure}
	\begin{subfigure}[b]{0.25\textwidth}
		\includegraphics[height=1\textwidth]{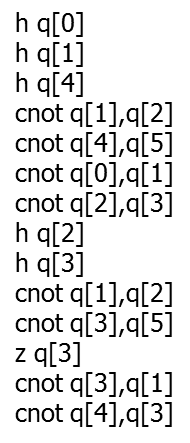}
		\caption{}
		\label{qasm}
	\end{subfigure}
\caption{An example circuit consisting of 6 qubits and 15 gates. (a) Its circuit description (top) and its virtual to physical qubits mapping (bottom) for the Surface-17 processor after initial placement; (b) Its cQASM representation without scheduling.}
\label{circuit_qasm}
\end{figure}

\begin{figure}[htb!]
\centering
	\begin{subfigure}[b]{0.2\textwidth}
		\includegraphics[height=1.5\textwidth]{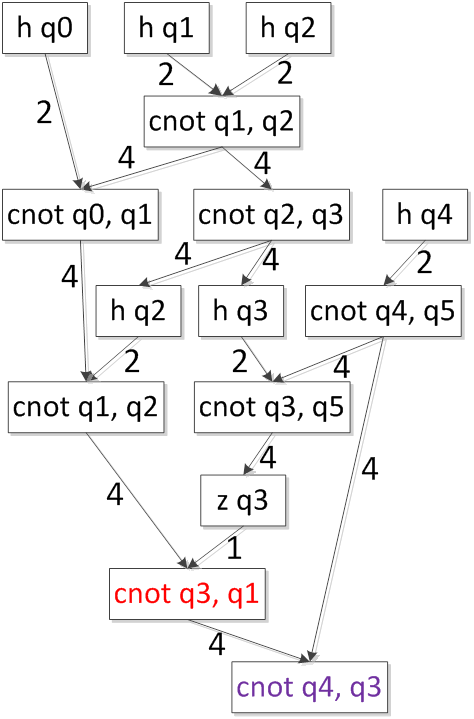}
		\caption{}
		\label{qodg}
	\end{subfigure}
	\begin{subfigure}[b]{0.25\textwidth}
		\includegraphics[height=1.3\textwidth]{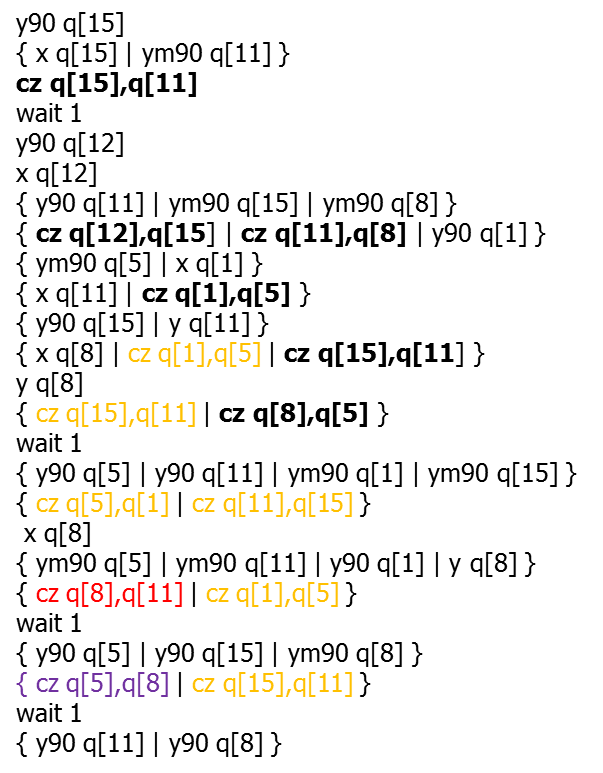}
		\caption{}
		\label{mappedqasm}
	\end{subfigure}
\caption{(a) The QODG of the circuit in Figure~\ref{circuit}. The red and purple boxed CNOTs have qubits that are not NN. (b) The parallel cQASM code of the mapped circuit, where each line represents one cycle and operations in the same line (inside one bracket) are scheduled to start at the same cycle. The CZ gates in bold are already nearest neighbouring. Movement operations (added two-qubit gates are in yellow) are inserted to perform the CZ gates in red and purple.}
\label{mappedcircuit}
\end{figure}

Similar to classical scheduling, a Quantum Operation Dependency Graph (QODG) $G(V_{G}, E_{G})$ is constructed from the QASM representation of a quantum circuit, in which each operation is denoted by a node $v_i \in V_{G}$, and the data dependency between two operations $v_i$ and $v_j$ is represented by a directed edge $e(v_i, v_j) \in E_{G}$ with weight $w_i$ that represents the duration of operation $v_i$. 
Pseudo source and sink nodes are added to the start and end to simplify starting and stopping iteration over the graph. 
The QODG of the circuit in Figure \ref{circuit} is shown in Figure \ref{qodg}.
In previous works that do not consider gate duration, only directed graphs are constructed, which cannot be directly applied to this work.

\subsection{Formulation of resource constraints} 
Furthermore, the scheduler also needs to adhere to the parallelism restrictions imposed by the shared classical control electronics as described in Section~\ref{sec:constraints}.
% , and this is achieved by the last module in Qmap, the RC-Scheduler.
In this work, these classical control constraints are treated as resource constraints in an otherwise conventional critical path list-scheduler implementation~\cite{kelley1959critical}. 
A so-called machine state $S$ is defined to describe the occupation status of each resource $r_{i}\in R$, where $R$ represents the set of all resources in $T_{g1}, T_{gm}, T_{g2f}$, and $T_{g2d}$.
The constraints for \textit{single-qubit gates} and \textit{measurement} are implemented by using $\left | T_{g1} \right |$ and $\left | T_{gm} \right |$ resource states, respectively. 
To support the \textit{two-qubit gates} constraint, there is a resource state for each connection (to constrain mutual CZ concurrency) and a resource state per qubit (to constrain CZ versus single-qubit gate concurrency).
Specifically, a resource state consists of two elements: the operation type that is using this resource and the occupation period which is described by a pair of cycle time ($[t_{0}, t_{1})$), representing the first cycle that it is occupied and the first cycle that it is free again, respectively.
If an operation $v$ is scheduled at cycle $t_{0}$ ($v.cycle=t_{0}$), then all the resources for performing $v$ ($v.resources$) will be occupied till (and not including) $t_{1}=t_{0}+v.duration$ ($v.duration$ is the duration of $v$).

\begin{algorithm}[H]
   \begin{algorithmic}[1]
        \Require Non-scheduled circuit
        \Require Configuration file with gate durations and resource descriptions $R$
		\Ensure Scheduled circuit 
		\State Generate QODG $G(V_{G}, E_{G})$ from circuit
		\State Initialize $\forall v \in V_{G}: v.resources \subset R$ and $v.duration$
        \State $V_{m} \gets$ Unique pseudo source node
        \State $V_{av} \gets$ All available gates in $G(V_{G}-V_{m}, E_{G})$
        \State Initialize cycle $t \gets 0$ 
        \State Initialize machine-state $S \gets \forall  r\in R$ is free
        \While{$V_{av} \neq \varnothing $}
            \State $V_{r} \gets$ resource-free gates $\subset V_{av}$ based on $S$
            \If{$V_{r} \neq \varnothing $}
                \State $V_{c} \gets$ Most-critical gates $\subset V_{r}$ in $G(V_{G}-V_{m}, E_{G})$
                \State Select $v \in V_{c}$ which is first in the circuit
                \State Add $v$ to $V_{m}$
                \State $v.cycle \gets t$
                \State Update $S$ with $v.resources$ occupied at $[t, t+v.duration)$
                \State $V_{av} \gets$ All available gates in $G(V_{G}-V_{m}, E_{G})$
            \Else 
                \State $t \gets t+1$
            \EndIf
        \EndWhile
   \end{algorithmic}
   \caption{Forward Scheduling algorithm}
   \label{alg:scheduling}
\end{algorithm}

\subsection{Scheduling heuristic}
% Here we describe the forward scheduling algorithm that is at the base of both the router and the global scheduler.
Algorithm~\ref{alg:scheduling} shows the pseudo code of this algorithm, which schedules all the gates of a given circuit with respect to the resource constraints.
Its objective is to achieve the shortest circuit latency.

The heuristic maintains two sets of gates: $V_{m}$ holds the gates that have been scheduled, and $V_{av}$ includes the gates that are available for scheduling.
A gate $v$ is \textbf{available} when all predecessors $p$ of $v$ in $G$ have been scheduled, that is, $\forall p, p$ is in $V_{m}$. Furthermore, it maintains a machine-state $S$ consisting of all resource states as described above.

Algorithm~\ref{alg:scheduling} first generates a QODG for the input circuit and initializes $V_{m}$, $V_{av}$, and $S$ (lines 1-6).
% Then it starts the schedule loops until $V_{av}$ gets empty.
After finding all the available gates at current cycle $t$, it selects the ones that can be scheduled at cycle $t$ and collects them in $V_{r}$ (line 8).
A gate $v \in V_{av}$ can be scheduled at cycle $t$ only if it is \textbf{resource-free} at $t$: when its predecessors have finished execution, i.e., $\forall p \in V_{m}, p.cycle + p.duration \leqslant t$ (this data dependency constraint can be seen as qubit resource constraint); and when all resources in $v.resources$ are not occupied for all cycles in $[t, t+v.duration)$. 
The worst-case time complexity of this step is $O(\min(g,n)\cdot(n+\left | R\right |))$, $n$ and $g$ are the number of qubits and operations in the input circuit, respectively (in the worst case, gates on every qubit can be scheduled.).

If $V_{r}$ is not empty, the heuristic selects the first most-critical gate $v$ in this set (lines 9-11).
A most-critical gate in $V_{r}$ is the one that has the longest path to the pseudo sink node of the QODG $G$.
In this work, the length of the longest path is pre-computed for each node in $G$, which only takes linear time.
Then it adds this gate $v$ to $V_{m}$, assigns the current cycle attribute to $v.cycle$, updates $S$ by reserving all the resources of $v$ ($v.resources$) for its execution duration, and updates $V_{av}$ given that $v$ has been scheduled now and thus some more gates may have become available (lines 12-15).
In this case, cycle $t$ is not incremented because more gates may be scheduled in the same cycle. 

If $V_{r}$ is empty, the heuristic increments $t$ (line 17) and continues the schedule loop again until all the gates are scheduled, that is, $V_{av}$ is empty.
In the worst case, this loop needs to be repeated $O(L)$ times, $L$ is the multiplication of the total number of operations ($g$) in the given circuit and the longest gate duration in cycles.
Resource-constrained scheduling is NP-hard in the strong sense~\cite{blazewicz1983scheduling}. 
Previous works that are using exact optimization approaches or exhaustive search algorithms for scheduling~\cite{lye2015determining,siraichi2018qubit,venturelli2018compiling,Zulehner2018efficient} cannot be adapted to efficiently solve this problem.
In contrast, the proposed scheduling algorithm has reduced its complexity to at most 
\[O_{schedule}=O\left (\min(g,n)\cdot(n+\left | R\right |)\cdot g\right).\]

\section{Mapping quantum algorithms}
\label{sec:mapper}

Mapping means to transform the original hardware-agnostic quantum circuit that describes the quantum algorithm to an equivalent one that can be executed on the target quantum processor. 
To this purpose, the mapping process has to be aware of the constraints imposed by the physical implementation of the quantum processor. 
These include the set of primitive gates that is supported, the allowed qubit interactions that are determined by the processor topology, and the limited concurrency of multi-gate execution because of classical control constraints.
Mapping will likely increase the number of operations that are required to implement the given algorithm as well as the circuit latency/depth, decreasing the reliability of the algorithm. 
Efficient algorithms that can minimize this mapping overhead are then necessary, especially in NISQ processors where noise sets a limit on the maximum size of a computation that can be run successfully.

\subsection{Overview of the Qmap mapper}
\begin{figure}[tbh!]
\centering
\includegraphics[width=0.45\textwidth]{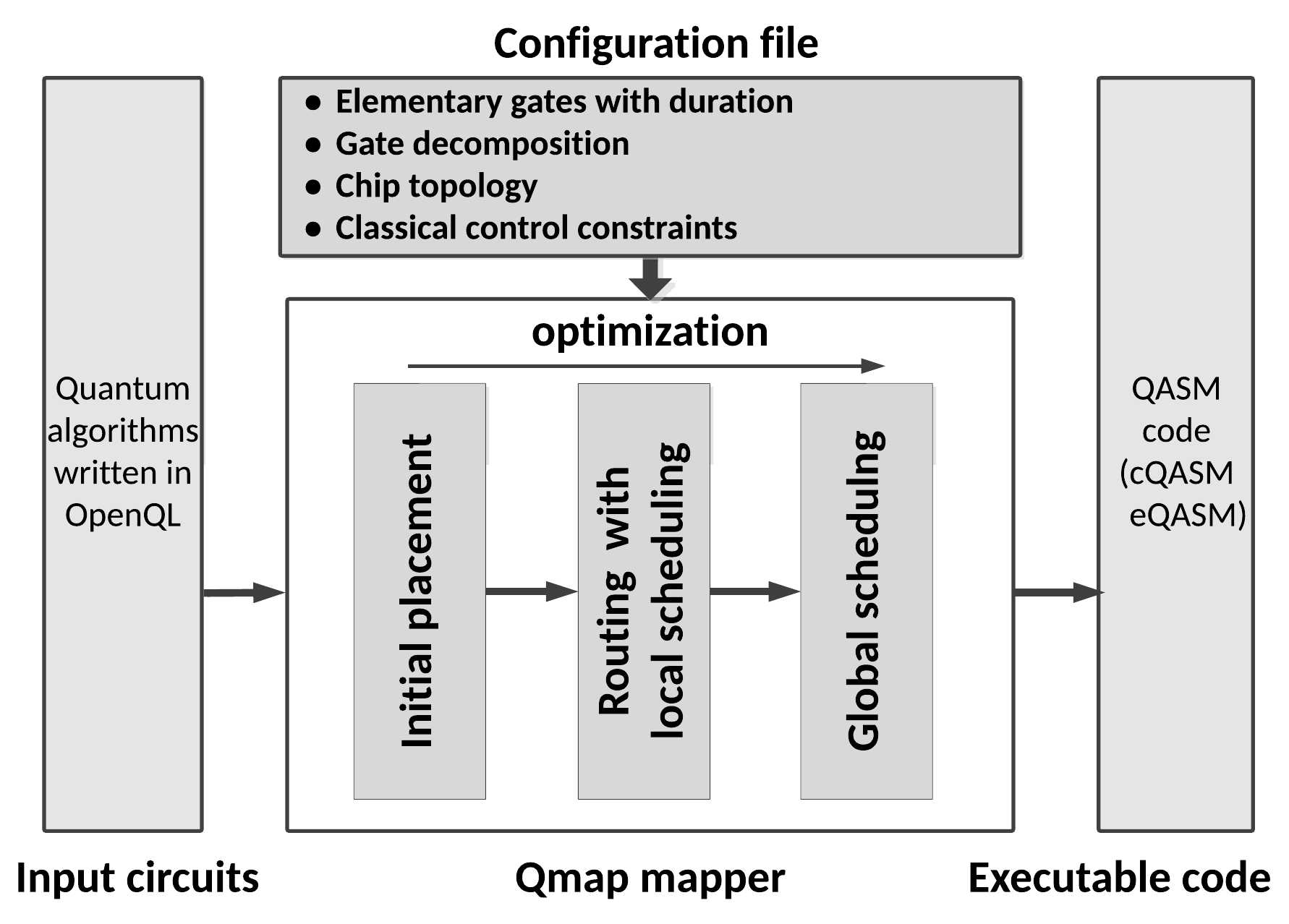}
\caption{Overview of the Qmap mapper embedded in the OpenQL compiler.}
\label{mapflow}
\end{figure}

The \textbf{Qmap} mapper developed in this work is embedded in the OpenQL compiler \cite{openql18qutech} and its design flow is shown in Figure \ref{mapflow}. 
The input of Qmap is a quantum circuit written in OpenQL (C++ or Python). 
The OpenQL compiler reads and parses it to a QASM-level intermediate representation. 
Qmap then performs the mapping and optimization of the quantum circuit based on the processor characteristics provided in a configuration file as described in the previous section. 
This approach allows Qmap to target different quantum devices by just changing the parameters in the configuration file. 
After mapping, QASM-like code is generated. 
Currently, the OpenQL compiler is capable of generating cQASM \cite{khammassi2018cqasm} that can be executed on the QX simulator \cite{khammassi2017qx} as well as eQASM \cite{fu2019eqasm}, a QASM-like executable code that can target the Surface-17 processor.
The generation of other QASM-like languages will be part of future extensions of the OpenQL compiler.
The modules of Qmap will be discussed in the rest of this section.

\subsection{Initial placement}  
It is preferable to place highly interacting qubits next to each other such that less movement operations will be added for performing two-qubit gates.
Similar to the placement approaches in~\cite{dousti2014squash}, the initial placement problem in this work is formulated as a quadratic assignment problem (QAP) and the objective is to minimize the
movement or communication overhead, which is modeled by the distance between interacting qubits minus 1.
Qmap tries to find an initial placement with minimum communication overhead by using the Integer Linear Programming (ILP) algorithm presented in \cite{lao2018mapping}. 
Such an initial placement implementation can only solve small-scale problems in reasonable time.
Even though for near-term implementations these numbers largely suffice, for large-scale circuits, one can either partition a large circuit into several smaller ones or apply heuristic algorithms to efficiently solve these models ~\cite{  dousti2013leqa}. 
Other works also solve this initial placement problem by using a Satisfiability Modulo Theories (SMT) solver \cite{murali2019full}. 

\subsection{Qubit Router}
\label{sec:router}
It is unlikely to find an initial placement in which all the qubit pairs that a two-qubit gate need to be performed on can be placed in neighboring positions. 
Therefore, qubits will have to be moved during computation.
For instance, based on the initial placement of qubits shown in Figure \ref{circuit}, the first 6 CNOT gates of the circuit can be performed directly as qubits are NN, but qubits in the last 2 CNOT gates will need to be routed to adjacent positions. 
Routing refers to the task of finding a series of movement operations that enables the execution of two-qubit gates on a given processor topology with low communication overhead. 
To do so, multiple routing paths are evaluated and one is selected based on various optimization criteria such as the number of added movement operations, increase of circuit depth, or decrease of circuit reliability ~\cite{qiskit, Zulehner2018efficient, siraichi2018qubit, finigan2018qubit,li2019tackling, tannu2019not, nishio2019extracting,rigetti, venturelli2018compiling, murali2019noise, murali2019full}. 
Afterwards, the corresponding movement operations are inserted. 

\begin{algorithm}[H]
   \begin{algorithmic}[1]
        \Require Non-routed circuit, VP-map $M$
        \Require Configuration file with topology and constraints
        \Ensure Routed circuit 
        \State Generate QODG $G(V_{G}, E_{G})$
        \State $V_{m} \gets$ Unique pseudo source node
        \State $V_{av} \gets$ All available gates in $G(V_{G}-V_{m}, E_{G})$
        \While{$V_{av} \neq \varnothing $}
            \State $V_{nn} \gets$ All single-qubit and NN two-qubit gates in $V_{av}$
            \If{$V_{nn} \neq \varnothing $}
                \State Select the first most-critical gate $v \in V_{nn}$ 
            \Else 
                \State $V_{c} \gets$ Most-critical gates $\subset V_{av}$ in $G(V_{G}-V_{m}, E_{G})$
                \State Select $v \in V_{c}$ which is first in the circuit
                \State Insert movement(s) for $v$
                \State Update $M$ 
            \EndIf 
            \State Map $v$ according to $M$
            \State Add $v$ to $V_{m}$
            \State $V_{av} \gets$ All available gates in $G(V_{G}-V_{m}, E_{G})$
        \EndWhile
   \end{algorithmic}
   \caption{Forward Routing algorithm}
   \label{alg:routing}
\end{algorithm}

\subsubsection{Routing heuristic}

In this work, after the ILP-based initial placement, a heuristic algorithm is used to perform this routing task.
It is a scheduler-based heuristic of which the objective is to minimize overall circuit latency.
Algorithm~\ref{alg:routing} shows the pseudo code of the proposed routing algorithm, which finds all two-qubit gates in which qubits are not nearest-neighbours and inserts the required movement operations to make them adjacent.
As mentioned in Section \ref{sec:constraints} we use SWAPs as well as MOVE operations for moving qubits.

The router algorithm starts by mapping the pseudo source node and then selecting all available gates ($V_{av}$) from the generated QODG (lines 1-3).
Then it finds all the single-qubit gates and the two-qubit gates of which qubits are NN from $V_{av}$, these gates are collected in $V_{nn}$ (line 5).
If $V_{nn}$ is not empty, then all gates in this set are mapped directly and a new set of available gates is computed (lines 6, 7, and 13-15). 
Mapping a (NN) gate implies replacing virtual qubit operands by their physical counterparts according to the VP-map table $M$ similar to the one shown in Figure \ref{circuit} and decomposing this gate to its primitives when the configuration specifies so. 

After that, only non-NN two-qubit gates remain in the available set. 
The router
selects the ones which are most critical in the remaining dependency graph $G$ since they have the highest likelihood to extend the circuit when mapped in an inefficient way or when delayed (line 9).
When there are several equally critical gates, the routing heuristic chooses the first one in the input circuit (line 10) and finds a set of movement operations to bring these two qubits to adjacent positions.
After the movement set selection, the router schedules the SWAP/MOVE operations into the circuit (line 11), updates the VP-map (line 12), recomputes the set of available gates (line 15), and runs the routing heuristic until all the gates are mapped.

\subsubsection{Movement set selection}
For finding a set of movement operations for a non-NN two-qubit gate, all shortest paths between these two qubits are considered. 
During Qmap initialization time, the distance (i.e. the length of the shortest path) between each pair of qubits has been computed using the Floyd-Warshall algorithm. 
Finding all shortest paths between qubits at mapping-time is done by a breadth-first search (BFS), that is, selecting only path extensions which decrease the distance between the qubits. 
For each shortest path, there may exist several movement sets since qubits can meet in any neighboring position within the path.
Note that all movement sets would lead to adding an equal minimum number of movements to the circuit.
In a $\sqrt{N}\times \sqrt{N}$ grid architecture, the total number of shortest paths between most remote two nodes $(q_{i}, q_{j})$ is $O(4^{\sqrt{N}})$ and the number of movement sets for each path is $(2\sqrt{N}-2)$.

In this work, a set of movement operations that minimally extends the circuit latency is selected and scheduled into the circuit.
As shown in Algorithm~\ref{alg:move}, this router evaluates all movement sets by looking back to the previously mapped gates (lines 1 and 2) and interleaving each set of movements with those gates using the proposed resource-constrained scheduling heuristic (Section~\ref{sec:schedule}) in an as-soon-as-possible (ASAP) policy (line 4). 
It selects the one(s) which minimally extend(s) the circuit latency (lines 6 and 7). 
When there are multiple minimal-cost sets, a random one is taken. 
The complexity of this routing strategy is $O(g\sqrt{n}4^{\sqrt{n}})\cdot O_{schedule}$. 

\begin{algorithm}[H]
  \begin{algorithmic}[1]
        \Require QODG $G(V_{G}, E_{G})$, gate $v$, VP-map $M$
        \Require Configuration file with topology and constraints
        \Ensure The set of movements for $v$
        \State $P \gets$ All shortest paths for $v$
        \State $MV_{P} \gets$ All possible sets of movements based on $P$
        \For{$mv_{j}$ in $MV_{P}$}
            \State Interleave $mv_{j}$ with previous gates (looking back)
            \State $L_{mv_{j}} \gets$ circuit's latency extension by $mv_{j}$
        \EndFor
        \If{$L_{mv_{i}} = \min(\bigcup_{j}{ L_{mv_{j}} })$}
            \State Select $mv_{i}$ as the set of movements, picking a random minimum one when there are more
        \EndIf
  \end{algorithmic}
  \caption{Movement selection algorithm}
  \label{alg:move}
\end{algorithm}

\subsection {Global scheduling}
After routing, the circuit adheres to the processor topology constraint for two-qubit interactions and has been scheduled in an as-soon-as-possible (ASAP) way.
The global scheduler reschedules the routed circuit to achieve the shortest circuit latency and the highest instruction-level parallelism.
It does this in an as-late-as-possible (ALAP) way to minimize the required life-time and thus the decoherence error of each qubit.  
The global scheduler employs a backward version of Algorithm~\ref{alg:scheduling}, i.e. it traverses the circuit starting from the sink node, working backwards through the circuit and decrementing $t$.

\subsection {Decomposition and optimization}
Starting from a quantum circuit described in cQASM format (see Figure \ref{circuit_qasm}), the circuit is also decomposed during mapping into one which only contains the \textit{primitive gates} specified in the configuration file, on top of adherence to the other constraints.
A circuit optimization module is also implemented to reduce the number of gates, e.g., two consecutive $X$ gates can cancel each other out.

The decomposition and optimization can be done at every step of the mapping procedure, i.e. before, during, and after routing.
Qmap reduces sequences of single qubit gates to their minimally required sequence both before and after routing.
Whether decomposition is applied at a mapping step is specified in the configuration file. 
The implementation of the QODG represents the commutability of not only all gates with disjoint qubit operands but also the known two-qubit operations CNOT and CZ with overlapping operands, and optimizes their order during both routing and global scheduling. 

The mapped version of the circuit in Figure~\ref{circuit} by using the Qmap mapper is shown in Figure \ref{mappedqasm}.
It is described in cQASM code with precise timing information, that is, which operations can be issued at each cycle. 
The output circuit can also be represented by eQASM code~\cite{fu2019eqasm} that can be directly read by the quantum microarchitecture in~\cite{fu2017experimental}.

\section{Qmap Evaluation}
\label{sec:results}

\begin{table*}[tbh!]
\centering
\caption{The characteristics of the input benchmarks including the number of qubits, the total number of gates, the number of two-qubit gates (CNOTs), its circuit depth and its circuit latency in cycles (20 ns per cycle).}
\label{tbl:bench}
\small
\resizebox{0.8\textwidth}{!}{
\begin{tabular}{|l|l|l|l|l|l|l|l|l|l|l|l|l|}
\cline{1-6} \cline{8-13}
Benchmarks         & Qubits & Gates & CNOTs & Depth & Latency & \multirow{29}{*}{} & Benchmarks      & Qubits & Gates & CNOTs & Depth & Latency \\ \cline{1-6} \cline{8-13} 
alu\_bdd\_288      & 7      & 84    & 38    & 48    & 169     &                    & sym9\_146       & 12     & 328   & 148   & 127   & 450     \\ \cline{1-6} \cline{8-13} 
alu\_v0\_27        & 5      & 36    & 17    & 21    & 72      &                    & sys6\_v0\_111   & 10     & 215   & 98    & 74    & 266     \\ \cline{1-6} \cline{8-13} 
benstein\_vazirani & 16     & 35    & 1     & 5     & 40      &                    & vbeAdder\_2b    & 7      & 210   & 42    & 52    & 116     \\ \cline{1-6} \cline{8-13} 
4gt12\_v1\_89      & 6      & 228   & 100   & 130   & 448     &                    & wim\_266        & 11     & 986   & 427   & 514   & 1788    \\ \cline{1-6} \cline{8-13} 
4gt4\_v0\_72       & 6      & 258   & 113   & 137   & 478     &                    & xor5\_254       & 6      & 7     & 5     & 2     & 5       \\ \cline{1-6} \cline{8-13} 
4mod5\_bdd\_287    & 7      & 70    & 31    & 40    & 140     &                    & z4\_268         & 11     & 3073  & 1343  & 1643  & 5688    \\ \cline{1-6} \cline{8-13} 
cm42a\_207         & 14     & 1776  & 771   & 940   & 3249    &                    & adr4\_197       & 13     & 3439  & 1498  & 1839  & 6377    \\ \cline{1-6} \cline{8-13} 
cnt3\_5\_180       & 16     & 485   & 215   & 207   & 729     &                    & 9symml\_195     & 11     & 34881 & 15232 & 19235 & 66303   \\ \cline{1-6} \cline{8-13} 
cuccaroAdder\_1b   & 4      & 73    & 17    & 25    & 58      &                    & clip\_206       & 14     & 33827 & 14772 & 17879 & 61786   \\ \cline{1-6} \cline{8-13} 
cuccaroMultiply    & 6      & 176   & 32    & 55    & 133     &                    & cm152a\_212     & 12     & 1221  & 532   & 684   & 2366    \\ \cline{1-6} \cline{8-13} 
decod24\_bdd\_294  & 6      & 73    & 32    & 40    & 143     &                    & cm85a\_209      & 14     & 11414 & 4986  & 6374  & 21967   \\ \cline{1-6} \cline{8-13} 
decod24\_enable    & 6      & 338   & 149   & 190   & 669     &                    & co14\_215       & 15     & 17936 & 7840  & 8570  & 29608   \\ \cline{1-6} \cline{8-13} 
graycode6\_47      & 6      & 5     & 5     & 5     & 20      &                    & cycle10\_2\_110 & 12     & 6050  & 2648  & 3384  & 11692   \\ \cline{1-6} \cline{8-13} 
ham3\_102          & 3      & 20    & 11    & 11    & 41      &                    & dc1\_220        & 11     & 1914  & 833   & 1038  & 3597    \\ \cline{1-6} \cline{8-13} 
miller\_11         & 3      & 50    & 23    & 29    & 105     &                    & dc2\_222        & 15     & 9462  & 4131  & 5242  & 18097   \\ \cline{1-6} \cline{8-13} 
mini\_alu\_167     & 5      & 288   & 126   & 162   & 564     &                    & dist\_223       & 13     & 38046 & 16624 & 19693 & 68111   \\ \cline{1-6} \cline{8-13} 
mod5adder\_127     & 6      & 555   & 239   & 302   & 1048    &                    & ham15\_107      & 15     & 8763  & 3858  & 4793  & 16607   \\ \cline{1-6} \cline{8-13} 
mod8\_10\_177      & 6      & 440   & 196   & 248   & 872     &                    & life\_238       & 11     & 22445 & 9800  & 12511 & 43123   \\ \cline{1-6} \cline{8-13} 
one\_two\_three    & 5      & 70    & 32    & 40    & 141     &                    & max46\_240      & 10     & 27126 & 11844 & 14257 & 49400   \\ \cline{1-6} \cline{8-13} 
rd32\_v0\_66       & 4      & 34    & 16    & 18    & 66      &                    & mini\_alu\_305  & 10     & 173   & 77    & 68    & 242     \\ \cline{1-6} \cline{8-13} 
rd53\_311          & 13     & 275   & 124   & 124   & 441     &                    & misex1\_241     & 15     & 4813  & 2100  & 2676  & 9240    \\ \cline{1-6} \cline{8-13} 
rd73\_140          & 10     & 230   & 104   & 92    & 330     &                    & pm1\_249        & 14     & 1776  & 771   & 940   & 3249    \\ \cline{1-6} \cline{8-13} 
rd84\_142          & 15     & 343   & 154   & 110   & 394     &                    & radd\_250       & 13     & 3213  & 1405  & 1778  & 6163    \\ \cline{1-6} \cline{8-13} 
sf\_274            & 6      & 781   & 336   & 436   & 1516    &                    & root\_255       & 13     & 17159 & 7493  & 8835  & 30575   \\ \cline{1-6} \cline{8-13} 
shor\_15           & 11     & 4792  & 1788  & 2268  & 7731    &                    & sqn\_258        & 10     & 10223 & 4459  & 5458  & 18955   \\ \cline{1-6} \cline{8-13} 
sqrt8\_260         & 12     & 3009  & 1314  & 1659  & 5740    &                    & square\_root\_7 & 15     & 7630  & 3089  & 3830  & 13049   \\ \cline{1-6} \cline{8-13} 
squar5\_261        & 13     & 1993  & 869   & 1048  & 3644    &                    & sym10\_262      & 12     & 64283 & 28084 & 35572 & 122564  \\ \cline{1-6} \cline{8-13} 
sym6\_145          & 7      & 3888  & 1701  & 2187  & 7615    &                    & sym9\_148       & 10     & 21504 & 9408  & 12087 & 41641   \\ \cline{1-6} \cline{8-13} 
\end{tabular}
}
\end{table*}

In this section, we evaluate \textit{Qmap} by mapping a set of benchmarks from RevLib~\cite{wille2008revlib} and QLib~\cite{lin2014qlib} on the superconducting processor Surface-17 that has a distance-3 surface-code topology~\cite{versluis2017scalable}.
All the hardware constraints discussed in Section~\ref{sec:constraints}, including the primitive gates with their real gate duration, the topology and the electronic control constraints are taken into account.
The mapping experiments are executed on a server with 2 Intel Xeon E5-2683 CPUs (56 logical cores) and 377GB memory. The Operating System is CentOS 7.5 with Linux kernel version 3.10 and GCC version 4.8.5.

\subsection{Benchmarks}

The circuit characteristics of the used benchmarks are shown in Table~\ref{tbl:bench}.
All circuits have been decomposed into ones which only consist of gates from the universal set \{Pauli, $S, S^{\dagger}, T, T^{\dagger}, H$, CNOT\}.
In these benchmarks, the number of qubits varies from 3 to 16, the number of gates goes from 5 to 64283, and the percentage of CNOT gates varies from 2.8\% to 100\%.
Moreover, the minimum circuit depth and the minimum circuit latency are also included, ranging from 2 to 35572 time-steps and from 5 to 12256 cycles (using the gate duration of Surface-17), respectively. Note that these numbers are meant to characterize the algorithms without considering the processor topology and classical control constraints.

The latter two parameters are defined as follows:

\textit{Circuit depth} is the length of the circuit. It is equivalent to the total number of time-steps for executing the circuit assuming each of the gates takes one time-step.

\textit{Circuit latency} refers to the execution time of the circuit considering the real gate duration. Latency and gate duration are expressed in cycles. In this paper, we assume that a cycle takes 20 nanoseconds. 

In order to generate quantum circuits which are executable on real processors, extra movement operations need to be added and gate parallelism will be compromised.
Other parameters after mapping these benchmarks to the Surface-17 processor are obtained, such as the number of inserted SWAP and MOVE operations and the CPU time the mapping process takes.
We analyze the impact of the mapping procedure in terms of number of gates and circuit latency for Surface-17.
The mapping overhead is calculated by $(X_{o}-X_{in})/X_{in}$, where $X_{in}$ and $X_{o}$ represent the values of the same circuit characteristic before and after mapping, respectively.

\begin{table*}[htb!]
\centering
\caption{The main differences of the Trivial, MinPath, and Qmap mappers.
$n$ and $g$ are the number of qubits and gates in an input circuit, respectively.}
\label{tbl:mapper_difference}
\resizebox{0.98\textwidth}{!}{
\begin{tabular}{|c|c|c|c|c|c|c|c|c|c|}
\hline
\multirow{2}{*}{} & \multirow{2}{*}{\begin{tabular}[c]{@{}c@{}}Circuit \\ optimization\end{tabular}} & \multirow{2}{*}{\begin{tabular}[c]{@{}c@{}}ILP-based \\ placement\end{tabular}} & \multicolumn{7}{c|}{Routing}                                                                                                                                                       \\ \cline{4-10} 
                  &                                                                                  &                                                                                         &
                  \begin{tabular}[c]{@{}c@{}}Smart gate \\ selection\end{tabular} &\begin{tabular}[c]{@{}c@{}}Shortest \\ path\end{tabular} & \begin{tabular}[c]{@{}c@{}}MOVE\\  operation\end{tabular} &
                  \begin{tabular}[c]{@{}c@{}}Multiple\\ movement sets\end{tabular} &\begin{tabular}[c]{@{}c@{}}Minimize\\  latency\end{tabular}
                  & \begin{tabular}[c]{@{}c@{}}wrt. Classical\\  controls\end{tabular}
                 & \begin{tabular}[c]{@{}c@{}}Complexity\end{tabular}\\
                  \hline
Trivial           & No                                                                               & No       &No                                                                              & Yes                                                      & No                                                        & No         & No        & No   &$O(g)$                                      \\ \hline
MinPath           & Yes                                                                              & Yes                                                                             & Yes        & Yes                                                      & Yes       &Yes                                                & No                          & No                 &$O(g\sqrt{n}4^{\sqrt{n}})$               \\ \hline
Qmap              & Yes                                                                              & Yes                                                                             & Yes        & Yes                                                      & Yes       & Yes                                                & Yes                         & Yes               &$O(g\sqrt{n}4^{\sqrt{n}})\cdot O_{schedule}$               \\ \hline
\end{tabular}
}
\end{table*}

\subsection{Prior mapping strategies}
As mentioned previously, the routing algorithms in most of prior works \cite{yazdani2013quantum, lye2015determining, wille2016look,farghadan2017quantum, herbert2018using, qiskit, Zulehner2018efficient, siraichi2018qubit, finigan2018qubit, li2019tackling, tannu2019not, nishio2019extracting,cowtan2019qubit, rigetti} optimise the number of operations, that is, the number of added SWAP gates.
They do not take actual gate duration and classical control limitations into account.
Their output circuits need to be further scheduled by a low-level hardware unit like OpenPulse~\cite{mckay2018qiskit} such that they can be correctly executed with precise timing.
In this work, we also implement such a mapping procedure called \textbf{MinPath} mapper to compare with the proposed timing and resource-aware Qmap that considers gate duration and control constraints during routing to minimize the circuit latency.
MinPath uses the same initial placement approach as the Qmap mapper.
However, the router in MinPath randomly selects one of the movement sets along one of the shortest paths as described in Section~\ref{sec:router} without respecting to classical control constraints and without evaluating which set(s) will minimally extend circuit latency.
The complexity of the router in MinPath is $O(g \sqrt{n}4^{\sqrt{n}})$.
               
Furthermore, we also introduce a \textbf{Trivial} mapper that may not be able to map the circuit with minimal latency extension but its routing strategy has linear complexity ($O(g)$).
In the trivial mapping strategy, a naive initial placement is used in which qubits are just placed in their appearing order, no circuit optimization is performed.
For the router in the trivial mapper, the gates in the input circuit are mapped in the order as they appear in the circuit, i.e. by-passing the QODG. 
For performing a non-NN two-qubit gate, it simply selects the first shortest path that is found. 
Moreover, only a single set of movement operations is generated for the chosen path, the set moving the control qubit adjacent to the target qubit. 
In addition, only SWAP gates are generated for moving qubits.
After routing, the proposed resource-constrained scheduling will be applied.
By contrast, the MinPath and Qmap mappers use the ILP-based initial placement, enable circuit optimization, and can insert both SWAP and MOVE gates.

The main differences of these three mapping strategies are summarized in Table~\ref{tbl:mapper_difference}.
To provide gate sequences with precise timing and comply with the classical control constraints, the proposed resource-constrained scheduling is also performed after routing procedure of the Trivial and MinPath mappers.

\subsection{Overhead comparison of different mappers}
Table~\ref{tbl:maptosc} shows the results of mapping benchmark circuits to the Surface-17 superconducting processor using three different mapping strategies: Trivial, MinPath, and Qmap. 
In this paper, the mapper is set to only find an ILP-based initial placement for the first ten two-qubit gates in any given circuit and computation time is limited to 10 minutes and is not included in the final CPU time.
For each benchmark circuit, the mapping procedure is executed for five times and the one with minimum overhead is reported.

Compared to the circuit characteristics before mapping (Table~\ref{tbl:bench}), no matter which strategy is applied, the mapping procedure results in high overhead for most of the benchmarks as shown in Table~\ref{tbl:maptosc}.
The only exceptions are the `benstein$\_$v' and `graycode6\_47' circuits, because some operations in these circuits can be canceled out by the optimization module in the mapper, decreasing their circuit sizes. 
When the trivial mapper is used, the mapping procedure leads to a high overhead in both circuit latency and total number of gates that ranges from $50\%$ (`graycode6$\_$47') to $ 1160\%$ (`xor5$\_$254') and from $122.9\%$ (`wim$\_$266') to  $800\%$ (`xor5$\_$254'), respectively.
The MinPath mapper results in an increase of the circuit latency and the total number of gates that goes from $38.9\%$ (`alu\_v0\_27') to $ 260\%$ (`xor5\_254')) and from $26.0\%$ (`cuccaroAdder\_1b') to $373.4\%$ (`rd84\_142'), respectively.
Finally, the proposed Qmap mapper increases the circuit latency and the total number of gates from $32.4\%$ (`miller\_11') to $ 260\%$ (`xor5\_254') and from $20.7\%$ (`cuccaroAdder\_1b') to $78.1\%$ (`rd32\_v0'), respectively.

\begin{figure*}[tbh!]
    \centering
    \includegraphics[width=\textwidth]{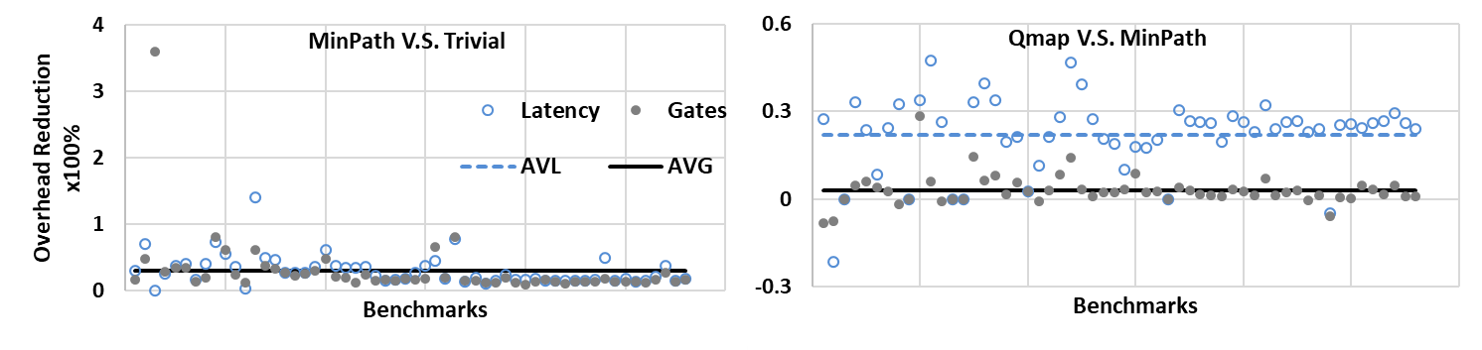}
    \caption{Comparison of three different mapping strategies. Overhead reduction (left) when comparing the MinPath mapper to the trivial mapper and (right) when comparing Qmap to MinPath.
    Benchmarks are in the horizontal axis and listed in their appearing order in Table~\ref{tbl:bench}.}
    \label{compare_router}
\end{figure*}

Furthermore, we compare the resulted overhead of these three mapping strategies as shown in Figure~\ref{compare_router}.
The trivial mapper leads to the highest mapping overhead as less optimization is performed.
Compared to the trivial strategy, the MinPath mapper can reduce the latency overhead and gate overhead up to $140\%$ (`gray6\_47') and $360\%$ (`benstein\_vazirani'), respectively.
The average latency (AVL) reduction and average gate (AVG) reduction are $30\%$ and $30.2\%$, respectively.
Moreover, the proposed Qmap mapper has lower or equal overhead than the MinPath mapper in terms of both circuit latency and number of gates for $96.4\%$ and $87.5\%$ of the benchmarks, respectively.
More specifically, Qmap can reduce the latency overhead up to $47.3\%$ (`decod24\_b') and decrease the gate overhead up to $28.6\%$ (`cuccaroMultiply') compared to the MinPath mapping strategy.
The average latency (AVL) reduction and average gate (AVG) reduction are $22\%$ and $3.0\%$, respectively.
This is because the router in the MinPath mapper only considers the qubit connectivity limitation and minimizes the number of operations, that is, it randomly selects a movement set that has minimum number of operations to move qubits to be neighbours. 
The gate duration and classical constraints will only be taken into account by a later module (such as the global scheduler in this work and the OpenPulse in IBM Qiskit~\cite{qiskit}).
In comparison, the router in Qmap uses the proposed resource-constrained scheduling approach as base and evaluates more minimum-weight movement sets to select one which minimally extends the circuit latency (Section~\ref{sec:mapper}).

\subsection{Scalability and runtime}
As discussed in Section~\ref{sec:mapper}, the complexity of the proposed resource-constrained scheduling heuristic in the worst case is still polynomial, making it applicable to large-scale quantum circuits.
The complexity of the routing heuristic is polynomial in terms of the number of gates but scales sub-exponentially with the number of qubits in a given circuit when using the Qmap and MinPath strategies.

We have tested three mapping strategies (Trivial, MinPath, Qmap) for different sizes of benchmarks, in which the number of qubits ranges from 3 to 16 and the two-qubit gate number from 5 to 62483.
The runtime (in seconds) that different mappers requires for mapping each benchmark on the Surface-17 processor can be found in Table~\ref{tbl:maptosc}, which is measured by the CPU time that the entire mapping procedure takes, excluding the time the ILP-based initial placement takes.
As expected, the mapper that performs more optimizations and evaluates more movement sets has a longer runtime. 
In this case, the trivial mapper has the shortest execution time whereas the Qmap takes the longest time.
For example, when mapping the largest benchmark `sym10\_262' with 62483 gates onto the Surface-17 processor, the trivial and the Qmap mappers take $ 72.8$ seconds and $9083.4$ seconds, respectively.

Based on the complexity analysis and the experimental results, we can conclude that Qmap is scalable in terms of large number of gates.
However, our experiments only use benchmarks which have less 20 qubits. 
Therefore, its scalability with the number of qubits needs to be further investigated.
Moreover, one may need to make a compromise between mapping performance and runtime for large-scale benchmarks.

\begin{table*}[htb!]
\centering
\caption{The results of mapping quantum benchmarks to the Surface-17 processor, including the total number of gates and the number of two-qubit gates (CZs) in the mapped output circuits, the circuit latency in cycles (20 ns per cycle), the numbers of inserted SWAP (SWs) and MOVE (MVs) operations, and the CPU time that mapping takes in seconds.}
\label{tbl:maptosc}
\small
\resizebox{1\textwidth}{!}{
\begin{tabular}{|l|l|l|l|l|l|l|l|l|l|l|l|l|l|l|l|l|l|l|}
\hline
\multirow{2}{*}{Benchmarks} & \multicolumn{6}{c|}{The Trivial mapper}               & \multicolumn{6}{c|}{The MinPath mapper}                 & \multicolumn{6}{c|}{The Qmap mapper}          \\ \cline{2-19} 
                            & Latency & Gates  & CZs   & SWs & MVs & Time       & Latency & Gates  & CZs   & SWs & MVs & Time      & Latency & Gates  & CZs   & SWS & MVs & Time      \\ \hline
alu\_bdd\_288               & 335     & 393    & 113   & 25    & 0     & 0.06365  & 286     & 341    & 100   & 16    & 7     & 1.7313    & 254     & 362    & 109   & 15    & 13    & 1.77362   \\ \hline
alu\_v0\_27                 & 166     & 188    & 56    & 13    & 0     & 0.0353  & 100     & 116    & 30    & 3     & 2     & 4.20968   & 106     & 122    & 34    & 3     & 4     & 4.13529   \\ \hline
benstein\_vazirani          & 36      & 45     & 10    & 3     & 0     & 0.01667  & 36      & 9      & 1     & 0     & 0     & 0.01051 & 36      & 9      & 1     & 0     & 0     & 0.01135 \\ \hline
4gt12\_v1\_89               & 931     & 1191   & 346   & 82    & 0     & 0.19592   & 811     & 917    & 270   & 54    & 4     & 25.6367   & 690     & 886    & 259   & 51    & 3     & 26.0342   \\ \hline
4gt4\_v0\_72                & 1124    & 1416   & 413   & 100   & 0     & 0.2555   & 884     & 1018   & 296   & 55    & 9     & 4.40794   & 788     & 973    & 273   & 52    & 2     & 4.628     \\ \hline
4mod5\_bdd\_287             & 298     & 339    & 100   & 23    & 0     & 0.07120  & 234     & 247    & 71    & 10    & 5     & 18.7469   & 226     & 240    & 69    & 10    & 4     & 19.5225   \\ \hline
cm42a\_207                  & 7167    & 8782   & 2532  & 587   & 0     & 1.42467    & 6499    & 7887   & 2352  & 517   & 15    & 611.534   & 5713    & 7724   & 2301  & 494   & 24    & 535.889   \\ \hline
cnt3\_5\_180                & 1985    & 2491   & 725   & 170   & 0     & 0.38054   & 1480    & 2103   & 623   & 136   & 0     & 25.0301   & 1236    & 2132   & 641   & 142   & 0     & 25.6028   \\ \hline
cuccaroAdder\_1b               & 175     & 171    & 50    & 11    & 0     & 0.03036  & 90      & 92     & 23    & 0     & 3     & 0.2521  & 90      & 92     & 23    & 0     & 3     & 0.28906  \\ \hline
cuccaroMultiply             & 417     & 427    & 122   & 30    & 0     & 0.06122  & 260     & 274    & 74    & 10    & 6     & 2.05933   & 217     & 246    & 64    & 6     & 7     & 2.09601   \\ \hline
decod24\_bdd                & 315     & 375    & 110   & 26    & 0     & 0.06353  & 253     & 301    & 90    & 14    & 8     & 1.38109   & 201     & 287    & 83    & 15    & 3     & 1.46449   \\ \hline
decod24\_enable             & 1342    & 1607   & 467   & 106   & 0     & 0.23441   & 1324    & 1464   & 434   & 95    & 0     & 28.704    & 1151    & 1474   & 434   & 95    & 0     & 28.7617   \\ \hline
graycode6\_47              & 30      & 31     & 11    & 2     & 0     & 0.00898 & 16      & 15     & 5     & 0     & 0     & 5.83973   & 16      & 15     & 5     & 0     & 0     & 5.8601    \\ \hline
ham3\_102                   & 79      & 87     & 26    & 5     & 0     & 0.01126  & 60      & 62     & 17    & 2     & 0     & 0.15724  & 60      & 62     & 17    & 2     & 0     & 0.22297   \\ \hline
miller\_11                  & 199     & 222    & 65    & 14    & 0     & 0.02856  & 156     & 166    & 46    & 3     & 7     & 0.19096  & 139     & 149    & 39    & 0     & 8     & 0.27471  \\ \hline
mini\_alu\_167              & 1144    & 1431   & 414   & 96    & 0     & 0.21319   & 985     & 1120   & 309   & 61    & 0     & 29.411   & 818     & 1068   & 294   & 56    & 0     & 28.5271   \\ \hline
mod5adder\_127              & 2229    & 2744   & 794   & 185   & 0     & 0.44677   & 1908    & 2240   & 645   & 130   & 8     & 7.0105   & 1618    & 2104   & 598   & 109   & 16    & 7.4544   \\ \hline
mod8\_10\_177               & 1819    & 2285   & 661   & 155   & 0     & 0.36898   & 1570    & 1808   & 530   & 102   & 14    & 2.26425   & 1434    & 1786   & 518   & 106   & 2     & 2.53567   \\ \hline
one\_two\_three             & 287     & 346    & 101   & 23    & 0     & 0.05446  & 235     & 263    & 76    & 12    & 4     & 6.18516   & 215     & 252    & 70    & 10    & 4     & 6.41456   \\ \hline
rd32\_v0\_66                & 168     & 184    & 55    & 13    & 0     & 0.02766  & 105     & 113    & 32    & 4     & 2     & 1.65692   & 104     & 111    & 31    & 1     & 6     & 1.71454   \\ \hline
rd53\_311                   & 1183    & 1514   & 448   & 108   & 0     & 0.24811   & 909     & 1249   & 370   & 78    & 6     & 0.32513  & 856     & 1257   & 375   & 81    & 4     & 0.67113  \\ \hline
rd73\_140                   & 970     & 1198   & 350   & 82    & 0     & 0.19047   & 751     & 1010   & 300   & 62    & 5     & 20.682    & 662     & 988    & 292   & 52    & 16    & 20.3441   \\ \hline
rd84\_142                   & 1385    & 1804   & 526   & 124   & 0     & 0.30129   & 1044    & 1624   & 481   & 109   & 0     & 20.7494   & 861     & 1516   & 448   & 98    & 0     & 21.1735   \\ \hline
sf\_274                     & 3351    & 3892   & 1137  & 267   & 0     & 0.67493   & 2705    & 3157   & 926   & 178   & 28    & 40.0639   & 2151    & 2822   & 818   & 104   & 85    & 41.2879   \\ \hline
shor\_15                    & 15082   & 19608  & 5472  & 1228  & 0     & 4.33023    & 13460   & 17464  & 5046  & 1028  & 87    & 2.45284   & 11217   & 17058  & 4924  & 982   & 95    & 14.6928   \\ \hline
sqrt8\_260                  & 12708   & 16131  & 4719  & 1135  & 0     & 3.49803    & 11626   & 14041  & 4231  & 953   & 29    & 4.12037   & 10020   & 13944  & 4216  & 956   & 17    & 13.2009   \\ \hline
squar5\_261                 & 7865    & 10178  & 2951  & 694   & 0     & 2.17597    & 7198    & 8922   & 2663  & 594   & 6     & 3.48788   & 6468    & 8764   & 2630  & 585   & 3     & 7.76352   \\ \hline
sym6\_145                   & 15466   & 19266  & 5583  & 1294  & 0     & 4.12125    & 14094   & 16427  & 4872  & 965   & 138   & 3.94839   & 12873   & 16145  & 4787  & 970   & 88    & 16.7757   \\ \hline
sym9\_146                   & 1250    & 1721   & 499   & 117   & 0     & 0.30173   & 1040    & 1493   & 447   & 93    & 10    & 21.1801   & 980     & 1456   & 431   & 91    & 5     & 21.6935   \\ \hline
sys6\_v0\_111               & 859     & 1142   & 338   & 80    & 0     & 0.24816   & 640     & 976    & 290   & 62    & 3     & 21.1563   & 573     & 909    & 267   & 49    & 11    & 21.1608   \\ \hline
vbeAdder\_2b                & 332     & 468    & 135   & 31    & 0     & 0.09799  & 236     & 300    & 79    & 9     & 5     & 0.16455   & 215     & 298    & 80    & 6     & 10    & 0.1938   \\ \hline
wim\_266                    & 3941    & 5084   & 1474  & 349   & 0     & 0.98658   & 3546    & 4289   & 1273  & 272   & 15    & 13.0377   & 3190    & 4203   & 1254  & 265   & 16    & 13.5583   \\ \hline
\textbf{xor5\_254}                   & 63      & 63     & 23    & 6     & 0     & 0.01135  & 18      & 18     & 8     & 1     & 0     & 29.7578   & 18      & 18     & 8     & 1     & 0     & 29.2384   \\ \hline
z4\_268                     & 12341   & 15792  & 4598  & 1085  & 0     & 3.19178    & 11463   & 13962  & 4178  & 905   & 60    & 818.036   & 9704    & 13537  & 4088  & 887   & 42    & 869.445   \\ \hline
adr4\_197                   & 14296   & 18110  & 5287  & 1263  & 0     & 3.38715    & 12772   & 15868  & 4780  & 1082  & 18    & 1.67818   & 11070   & 15496  & 4685  & 1021  & 62    & 10.924    \\ \hline
9symml\_195                 & 142144  & 182319 & 53224 & 12664 & 0     & 36.5722    & 134023  & 164219 & 49485 & 11167 & 376   & 16.642    & 116118  & 162001 & 49154 & 11282 & 38    & 2332.7    \\ \hline
clip\_206                   & 139948  & 180243 & 52809 & 12679 & 0     & 40.1273    & 128597  & 162421 & 49227 & 11379 & 159   & 17.44     & 111253  & 160880 & 49090 & 11268 & 257   & 2587.99   \\ \hline
cm152a\_212                 & 5166    & 6320   & 1834  & 434   & 0     & 1.3859     & 4508    & 5347   & 1586  & 346   & 8     & 0.66896  & 4086    & 5306   & 1591  & 353   & 0     & 3.53968   \\ \hline
cm85a\_209                  & 48394   & 61007  & 17886 & 4300  & 0     & 14.2237    & 44110   & 54845  & 16654 & 3832  & 86    & 6.49007   & 37839   & 53363  & 16224 & 3716  & 45    & 389.036   \\ \hline
co14\_215                   & 75821   & 99108  & 29218 & 7126  & 0     & 20.9755    & 68064   & 92308  & 28381 & 6837  & 15    & 10.8777   & 57968   & 90267  & 27787 & 6615  & 51    & 1044.04   \\ \hline
cycle10\_2\_110             & 25607   & 31630  & 9236  & 2196  & 0     & 7.12406    & 23070   & 28148  & 8460  & 1904  & 50    & 3.26144   & 20458   & 27897  & 8471  & 1899  & 63    & 106.071   \\ \hline
dc1\_220                    & 7740    & 9845   & 2867  & 678   & 0     & 2.62955    & 7116    & 8575   & 2574  & 567   & 20    & 1.45486   & 5979    & 8117   & 2444  & 481   & 84    & 8.51783   \\ \hline
dc2\_222                    & 39466   & 50396  & 14754 & 3541  & 0     & 12.5991    & 36113   & 44864  & 13547 & 3100  & 58    & 5.16826   & 31796   & 44379  & 13520 & 3077  & 79    & 268.637   \\ \hline
dist\_223                   & 156674  & 201426 & 58891 & 14089 & 0     & 40.4085    & 144079  & 183197 & 55613 & 12757 & 359   & 55.0312   & 124031  & 179639 & 54717 & 12599 & 148   & 3550.7    \\ \hline
ham15\_107                  & 36221   & 45826  & 13356 & 3166  & 0     & 10.1604    & 33368   & 40721  & 12257 & 2797  & 4     & 5.74947   & 28906   & 39762  & 12030 & 2704  & 30    & 193.48    \\ \hline
life\_238                   & 92286   & 117371 & 34238 & 8146  & 0     & 30.3595    & 85068   & 104447 & 31370 & 7134  & 84    & 14.0716   & 75462   & 104689 & 31920 & 7324  & 74    & 1364.49   \\ \hline
max46\_240                  & 111978  & 141438 & 41211 & 9789  & 0     & 35.6086    & 101798  & 125209 & 37631 & 8375  & 331   & 16.8426   & 89164   & 123895 & 37565 & 8217  & 535   & 1840.89   \\ \hline
mini\_alu\_305              & 767     & 862    & 254   & 59    & 0     & 0.16079   & 505     & 741    & 228   & 35    & 23    & 0.19804  & 518     & 775    & 242   & 41    & 21    & 0.40925  \\ \hline
misex1\_241                 & 19670   & 24793  & 7206  & 1702  & 0     & 5.75472    & 18143   & 22002  & 6577  & 1479  & 20    & 3.13745   & 15892   & 21883  & 6588  & 1480  & 24    & 53.4152   \\ \hline
pm1\_249                    & 7167    & 8782   & 2532  & 587   & 0     & 2.3615     & 6446    & 7793   & 2314  & 499   & 23    & 1.48028   & 5629    & 7774   & 2331  & 504   & 24    & 6.86838   \\ \hline
radd\_250                   & 13254   & 16700  & 4867  & 1154  & 0     & 4.11447    & 12291   & 14955  & 4516  & 979   & 87    & 2.43807   & 10798   & 14408  & 4363  & 948   & 57    & 24.0061   \\ \hline
root\_255                   & 71310   & 91873  & 26882 & 6463  & 0     & 20.7858    & 64948   & 82599  & 24991 & 5824  & 13    & 11.4199   & 55963   & 80542  & 24520 & 5627  & 73    & 844.114   \\ \hline
sqn\_258                    & 43328   & 53252  & 15529 & 3690  & 0     & 11.7106    & 38165   & 46370  & 13908 & 3019  & 196   & 6.55198   & 33010   & 45801  & 13815 & 2984  & 202   & 270.981   \\ \hline
square\_root\_7             & 35769   & 44042  & 12896 & 3269  & 0     & 10.6409    & 27419   & 34333  & 10274 & 2371  & 36    & 50077.7   & 23203   & 33088  & 9845  & 2184  & 102   & 46862.9   \\ \hline
sym10\_262                  & 269200  & 340622 & 99658 & 23858 & 0     & 72.7567    & 247750  & 305153 & 92270 & 21030 & 548   & 42.2372   & 215185  & 303141 & 92326 & 20986 & 642   & 9083.41   \\ \hline
sym9\_148                   & 87919   & 110393 & 32127 & 7573  & 0     & 27.4377    & 79881   & 95215  & 28378 & 6152  & 257   & 14.4444   & 70756   & 94656  & 28462 & 6182  & 254   & 800.226   \\ \hline
\end{tabular}
}
\end{table*}

\subsection{SWAPs versus MOVEs}

\begin{figure}[tbh!]
    \centering
    \includegraphics[width=0.5\textwidth]{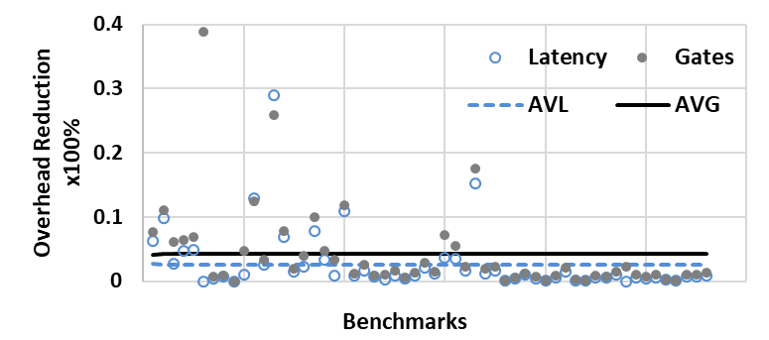}
    \caption{Reduction of mapping overhead when using MOVEs if possible compared to when only using SWAPs.
    Benchmarks are in the horizontal axis and listed in their appearing order in Table~\ref{tbl:bench}.
    The average latency (AVL) reduction and average gate (AVG) reduction are 2.76\% and 4.21\%, respectively.}
    \label{mv_sw}
\end{figure}

As mentioned in Section~\ref{sec:constraints}, a SWAP gate is implemented by three consecutive CNOT gates whereas a MOVE operation is implemented by two consecutive CNOT gates but requiring an ancilla qubit in the state $\ket{0}$.
Therefore, if there are available ancilla qubits (qubits that are not used for computation), then it is preferable to use MOVE operations rather than SWAP gates, which helps to reduce the mapping overhead.
In this section, we evaluate the benefit of using MOVE operations instead of only using SWAPs.
We map the benchmarks in Table~\ref{tbl:bench} onto the Surface-17 processor using the MinPath mapper. 
Different from the setups in Table~\ref{tbl:maptosc}, to have a fair comparison between using MOVEs if possible and only using SWAPs, in this case the native initial placement is applied and the first movement set is always selected.
With the same qubit overhead, the mapping with MOVEs can reduce the number of gates up to $38.9\%$ (`bestein\_vazirani') and the circuit latency up to $29\%$ (`graycode6\_47') compared to the mapping with only SWAPs as shown in Figure~\ref{mv_sw}.
The latency reduction and gate reduction are higher than 1\% for around 48.2\% and 64.3\% of the benchmarks, respectively.

\section{Related work}
\label{sec:review}
To achieve the shortest circuit latency and provide precise timing information to generate correct control signals, schedulers that consider actual gate duration should be developed.
Furthermore, the control electronic constraints that can be very restrictive especially when scaling-up quantum processors, should also be taken into account to allow valid execution of quantum applications.
As discussed previously, most of prior mapping works~\cite{ yazdani2013quantum,  lye2015determining, wille2016look,farghadan2017quantum, herbert2018using,qiskit, Zulehner2018efficient, siraichi2018qubit, finigan2018qubit, li2019tackling, tannu2019not, nishio2019extracting,cowtan2019qubit,rigetti} only focus on the primitive gate set and qubit connectivity constraints.
The output circuits from prior mappers need to be further scheduled with respect with the gate duration and classical control constraints, which is less optimal than the Qmap mapper as shown in Section~\ref{sec:results}.
Moreover, they all use SWAP operations for moving qubits when targeting superconducting quantum processors. 
In addition, so far no mapper has been developed for more scalable quantum processors such as the Surface-17 processor presented in \cite{versluis2017scalable, Intelwebsite}. 
Although this type of processors has been designed with the aim of building a large qubit array capable of performing fault-tolerant quantum computations based on surface code, it can be also used for running quantum algorithms in a near-term implementation.   

Many existing mapping algorithms~\cite{ yazdani2013quantum,  lye2015determining, wille2016look,farghadan2017quantum, herbert2018using,qiskit, Zulehner2018efficient, siraichi2018qubit, finigan2018qubit, li2019tackling, tannu2019not, nishio2019extracting,cowtan2019qubit,rigetti, venturelli2018compiling, booth2018comparing, venturelli2019quantum} and this paper use either the number of inserted movement operations or the circuit depth/latency as optimization metrics.
Although all these metrics affect the success probability of a quantum circuit, an analysis on which ones are more critical to be minimized is required.
Recent works \cite{finigan2018qubit, tannu2019not, nishio2019extracting, murali2019noise, murali2019full, linke2017experimental}
suggest to choose the routing path based on the fidelity of the two-qubit gates along the path as they are used to implement the movements (noise-aware mapper).
However, the reliability of a path is calculated by simply multiplying the fidelity of each gate without considering error propagation and decoherence, which makes this metric incomplete and not very accurate and it thus sometimes fails in predicting the most reliable route \cite{nishio2019extracting}. 
A more accurate metric that can well represent success probability and also can be easily used by the mapping procedure needs to be developed.

\section{Conclusion and Discussion}
\label{sec:conclusion}
In this work, we have presented a mapper called Qmap to make quantum circuits executable on the Surface-17 superconducting processor. 
It takes into account common processor constraints such as the primitive gate set and qubit connectivity, as well as actual gate duration and classical control electronic restrictions as it minimizes the circuit latency. 
Qmap has been embedded in the OpenQL compiler and consists of several modules, including qubit initial placement, resource-constrained list scheduling with polynomial complexity, qubit routing, and gate decomposition and optimization. 
The evaluation results show that the proposed timing and resource-aware mapper results in lower overhead in terms of both circuit latency and number of gates compared to the prior mapping strategy (MinPath) that minimizes the number of operations in the routing process and then reschedules the circuits with respect to the actual gate duration and classical control constraints.
However, the complexity of the routing algorithm in Qmap scales sub-exponentially with the number of qubits in the input circuit.
Future work can reduce its complexity by only evaluating the shortest paths where less qubits were, are or will be busy in the past, current, or coming cycles. 

Furthermore, Qmap can be applied to different processors by only changing their corresponding hardware characteristics in the configuration file.
We will investigate the performance of Qmap on other NISQ processors and compare it with prior works in the future.
In addition, more mapping metrics need to be investigated and included in the mapper. Note that what parameter(s) to optimise during the mapping might depend on the characteristics of the target quantum processor.
In addition, our mapping approach is based on the compilation of quantum circuits at the gate level.
Although it generates valid instructions with precise timing, they still need to be further translated into appropriate signals that control the qubits by the microarchitecture proposed in~\cite{fu2017experimental}.
A different approach is to directly compile quantum algorithms to control pulses~\cite{shi2019optimized}.
Further work will compare both solutions and investigate the trade-off of allocating mapping tasks to the compiler and the microarchitecture.

\begin{acknowledgments}
The authors would like to thank Xiang Fu and Adriaan Rol for useful discussions. 
The authors acknowledge support from the Intel Corporation. LLL also acknowledges funding from the China Scholarship Council.
\end{acknowledgments}
% \newpage
\bibliography{references}
\end{document}